\documentstyle[preprint,aps]{revtex}
\begin{document}
\draft
\date{\today}
\title{Excitons in insulating cuprates}
\author{M.E. Sim\'{o}n, A.A. Aligia, C.D. Batista, E. R.Gagliano, and F. Lema}
\address{Centro At\'omico Bariloche and Instituto Balseiro}
\address{Comisi\'on Nacional de Energ{\'\i}a At\'omica}
\address{8400 S.C. de Bariloche, Argentina.}
\maketitle

\begin{abstract}
We study the electronic excitations near the charge-transfer gap in
insulating CuO$_2$ planes, starting from a six-band model which includes $%
p_\pi $ and $d_{xy}$ orbitals and Cu-O nearest-neighbor repulsion $U_{pd}$.
While the low lying electronic excitations in the doped system are well
described by a modified $t-J$ model, the excitonic states of the insulator
include hybrid $d_{xy}-$ $p_\pi $ states of $A_{2g}$ symmetry. We also
obtain excitons of symmetries $B_{1g}$ and $E_u$, and eventually $A_{1g}$,
which can be explained within a one-band model. The results agree with
observed optical absorption and Raman excitations.
\end{abstract}


There is general consensus that the main electronic properties of
superconducting CuO$_2$ planes are well described by the three-band Hubbard
model  $H_{3b}$ \cite{var}  or effective one-band models derived from
it \cite{zha,jef,bel,hyb,hyb2,bat,sch,sim,sim2}. In particular, accurate 
low-energy reductions of $H_{3b}$ have been made using the cell-
perturbation method \cite{jef,bel,sim2} or fitting energy levels 
\cite{hyb,hyb2,bat} and the resulting generalized
$t-J$ or Hubbard models were used to explain many fundamental properties
of these materials, like superconductivity \cite{sup}, angle-resolved
photoemission \cite{bon}, x-ray absorption spectrum
in La$_{2-x}$Sr$_x$CuO$_4$ \cite{hyb2,che}, metal-insulator
transition as a function of the charge-transfer energy $\Delta $ \cite{sim}
and Raman scattering \cite{sha,chu}.

While the general features of optical absorption experiments \cite{fal,per}
have been explained using $H_{3b}$ \cite{lor}, recent Raman experiments
\cite{liu} in several insulating cuprates show excitations below the 
charge-transfer gap which cannot be explained with $H_{3b}$. This 
raises serious doubts on the validity of the very large amount of 
theoretical work ultimately based on this model. The most intense 
observed Raman transitions have symmetries $A_{2g}$ and $B_{1g}$ .
The former lies lower in energy, is more intense for the 
frequency  of the incident light used, and has been ascribed to
transitions  from $d_{x^2-y^2}$ to $d_{xy}$ or $p_{\pi}$ orbitals
\cite{liu}. The latter two orbitals are not contained in $H_{3b}$. The
$B_{1g}$ peak follows closely the optical absorption edge and its
origin is not clear so far.

 The purpose of this work is twofold. First, to explain within a
unifying  approach the $E_u$ exciton observed in optical absorption
experiments, and the main features of the Raman 
experiments in insulating systems. Second, to analyze the more 
fundamental issue of how to reconcile the $A_{2g}$ peak with our 
present theoretical understanding of the cuprates.
 We study a six-band model $H_{6b}$ obtained generalizing $%
H_{3b}$ to include the $d_{xy}$ and $p_\pi $ orbitals. The model is treated
within the cell-perturbation method \cite{jef,bel,sim}. This method has been
shown to give accurate results \cite{jef,sim} and at the present, we believe
that there are no accurate alternative methods to handle this particular
problem. We obtain in agreement with experiment that the observed
absorption $E_u$ peak and the Raman $B_{1g}$ peak lie very
close in energy, and (0.1-0.2) eV below the absorption edge. While these
peaks could be explained within an effective generalized one-band Hubbard
model, the Raman $A_{2g}$ peak corresponds to an excitation of a hole from a
(predominantly) $d_{x^2-y^2}$ orbital to a linear combination of $d_{xy}$
and $p_\pi $ holes with $b_{2g}$ symmetry. However, holes added to the 
ground state of the insulating system enter predominantly $2p_\sigma$
orbitals in agreement with experiment \cite{tak} and our previous
understanding of these systems.

The six-band Hamiltonian can be written as:

\begin{eqnarray}
H_{6b} &=&\sum_{i\sigma }\Delta _{xy}n_{ia\sigma }+\sum_{j\sigma }\left(
\Delta n_{jp\sigma }+\Delta _\pi n_{j\pi \sigma }\right) +U_d\sum_{i,\beta
\sigma \neq \ \beta ^{\prime }\sigma ^{\prime }}^{}n_{i\beta \sigma }n_{i\
\beta ^{\prime }\sigma ^{\prime }}+U_p\sum_{j,\alpha \sigma \neq \alpha
^{\prime }\sigma ^{\prime }}n_{j\alpha \sigma }n_{j\alpha ^{\prime }\sigma
^{\prime }}  \nonumber \\
&&+U_{pd}\sum_{i\delta \alpha \beta \sigma \sigma ^{\prime }}n_{i \beta \sigma
}n_{i+\delta \alpha \sigma ^{\prime }}+t_{pd}\sum_{i\delta \sigma }f(\delta
)(p_{i+\delta \sigma }^{\dagger }d_{i\sigma }+h.c.)+t_{\pi a}\sum_{i\delta
\sigma }g(\delta )(\pi _{i+\delta \sigma }^{\dagger }a_{i\sigma }+h.c.) 
\nonumber \\
&&+\sum_{j\gamma \sigma }[h(\gamma )(t_{pp}p_{j+\gamma \sigma }^{\dagger
}p_{j\sigma }+t_{\pi \pi }\pi _{j+\gamma \sigma }^{\dagger }\pi _{j\sigma
})-t_{p\pi }(\pi _{j+\gamma \sigma }^{\dagger }p_{j\sigma }+h.c.)]
\end{eqnarray}

\noindent with $n_{i\beta \sigma }=\beta _{i\sigma }^{\dagger }\beta
_{i\sigma }$, $\beta =d$ or $a$; $n_{j\alpha \sigma }=\alpha _{j\sigma
}^{\dagger }\alpha _{j\sigma }$, $\alpha =p$ or $\pi $; $d_{i\sigma
}^{\dagger }$, $a_{i\sigma }^{\dagger }$, $p_{j\sigma }^{\dagger }$,$\pi
_{j\sigma }^{\dagger }$ create a $3d_{x^2-y^2}$, $3d_{xy}$,$~2p_\sigma $ or $%
2p_\pi $ hole respectively with spin $\sigma $ at Cu site $i$ or O site $j$.
The indices\ $i+\delta \;(j+\gamma )$ label the four O atoms nearest to the
Cu (O) atom at site $i~(j)$. The functions $f(\delta )$, $g(\delta )$ and $%
h(\delta )$ describe the sign of the hoppings as functions of direction.

Most of the parameters of the model are known, at least approximately, from
constrained-local-density-functional\ calculations \cite{hyb,gra,ann}. The
parameters involving the $p_\pi $ orbitals are, according to the study
of Mattheis and Hamann\ \cite{mat}, $%
t_{p\pi }=0.45t_{pp},\;t_{\pi \pi }=0.48\;t_{pp}$, and\ $\Delta _\pi \cong
\Delta -1$eV. We take $t_{pd}\simeq 1.2eV$ as the unit of energy and the
following parameters  \cite{hyb,gra,ann,mat}: $\Delta =2.35,\;\Delta
_{xy}=1.35$, $\Delta _\pi =1.55,\;U_d=8.33,\;U_p=3.33,\;U_{pd}=0.833,\;t_{\pi
a}=0.65,$ $t_{pp}=0.5,\;t_{\pi \pi }=0.3,\;t_{p\pi }=t_{\pi \pi }.$

The large number of orbitals per unit cell and the highly correlated nature
of $H_{6b}$, make impossible at present an exact treatment of the model in a
large enough system . However, the methods we use are accurate enough to
explain the nature of the observed optical excitations. Our basic approach
is the cell-perturbation method \cite{jef,bel,sim}. For each Cu site\ $i$,
four orthogonal Wannier functions of the O $2p$ orbitals centered at that
site are defined (see Fig. 1). Two of them, of symmetry $b_{1g}$ (like the $%
3d_{x^2-y^2}$ orbital) and $a_{1g}$ (''non-bonding'') correspond to the $%
p_\sigma $ orbitals. The Wannier functions constructed with the\ $p_\pi $
orbitals have symmetry $a_{2g}$ and $b_{2g}$. The cell Hamiltonian $H_i$
contains all interactions and hoppings involving only the six orthogonal
Wannier functions centered at site $i$. For each cell, $H_i$ is diagonalized
exactly and the intercell part is written in terms of
the eigenstates of lowest energy of $H_i$. The advantage of this method is
that $H_i$ contains the highest energy in the problem ($U_d$)\ and most of
the $U_{pd}$ and hopping terms. For one hole in the cell, the ground state
of $H_i,~\;\left| ig\sigma \right\rangle $ has predominantly Cu character
and is a linear combination of both $b_{1g}$ states with spin $\sigma $ \cite
{jef,bel,sim}. The first excited state $\left| i\pi \sigma \right\rangle $
corresponds to one hole in a linear combination of the $d_{xy}$ and O\ $%
b_{2g}$ Wannier functions and lies 1.63 above $\left| ig\sigma \right\rangle 
$ for the parameters given before. For two holes, the ground state of $%
H_i,\;\left| ig2\right\rangle $ is a linear combination of the three
singlets constructed with both $b_{1g}$ orbitals, and \ corresponds to the
Zhang-Rice singlet (ZRS) of the three-band model \cite{wag} and one-band
models derived from it \cite{zha,jef,bel,sch,sim}. The first excited
state lying 0.99 above, denoted as $\left| i\pi 2\right\rangle $, is spin
degenerate and contains one $a_{2g}$ hole and one $b_{1g}$ hole of
predominantly Cu character. The hopping
$\left| i g 2\right\rangle \left| j g 1\right\rangle \rightarrow 
\left| i g 1\right\rangle \left| j \pi 2\right\rangle $ vanishes by
symmetry if $i$ and $j$ are nearest neighbors (NN). The larger hopping of this
kind is for next-NN and is $0.45 t_{p\pi}=0.13$, 
considerably smaller than the energy difference 0.99 and about a third of
the hopping between ZRS. Thus, added holes in the system form ZRS and
the amount of $p_\pi$ holes in the ground state of hole-doped systems is
very small for small doping. The same happens for $d_{xy}$ holes. 
For the explicit calculation of the exciton energies we neglect these 
small hoppings. 

The $\left| ig\sigma
\right\rangle ,$ $\left| ig2\right\rangle $ states are mapped
into a generalized one-band model $H_{1b}$ which includes NN, next-NN hopping and NN
repulsions which depend on the occupation of the sites involved \cite
{sch,sim}.
The NN repulsions are of order $U_{pd}/4.$ The movement of a vacant site (Cu$%
^{+}$ state) and that of a ZRS in the insulating system are described by
corresponding $\;t-t^{\prime }-J$ models with $J=0.12$ (near to the
experimental value) and the other parameters taken from $H_{1b}$: for the
movement of the ZRS (vacant site) we obtain $t_h=0.39,\;t_h^{\prime
}=0.034\;(t_e=0.34,\;t_e^{\prime }=0.005)$ in units of $t_{pd}.$ The problem
of the exciton energy is then approximated by that of an electron and a hole
moving freely in their respective quasiparticle bands, except at NN, where
they feel an attraction $V_1\sim 0.18$ and an on-site interaction determined
by the eigenstates of $H_i$. A similar approach has been used previously by
Belinicher {\it et al.} \cite{bel}. The quasiparticle band is taken of the
form \cite{mar,bon}: 
\begin{equation}
\epsilon _{ki}=\epsilon _{0i}+4t_{2i}\cos k_x\cos k_y+2t_{3i}(\cos 2k_x+\cos
2k_y)
\end{equation}
\noindent where $i=h(e)$ for holes (electrons). The parameters are
determined from $t,~t^{\prime }$ and $J$ by the self-consistent Born
approximation. We obtain $\epsilon _{0h}=-0.507,~t_{2h}=0.0085,~t_{3h}=0.017~~
$ ($\epsilon _{0e}=-0.507,~t_{2e}=0.021,~t_{3e}=0.015$). The resulting  
energy gap is $1.39$.

For total wave vector ${\bf K}=0$, the problem takes the form of a
one-particle one. For symmetry $A_{1g}$ we do not obtain a bound state
because $V_1$ is slightly smaller than the critical value (0.19) necessary
to obtain an exciton. For symmetries$~B_{1g}$ and $E_u$ the resulting
binding energies are 0.09 and 0.11 respectively. A scheme of the
quasiparticle bands and these excitonic energies is represented at the right
of Fig. 2. It can be shown that the binding energies of the $E_u$, $B_{1g}$
and $A_{1g}$ excitons are mainly related with the quasiparticle energies at ($%
\pi /2$, $\pi /2$), ($\pi $, 0) and (0, 0) respectively. The lower the
quasiparticle energy of the $t-t^{\prime }-J$ model at one of these points,
the higher the binding energy of the respective mode.

The amplitude of the excitonic wave functions for each irreducible
representation is shown in Fig. 3. For symmetry $A_{1g}$, there is a small
probability of finding the electron and hole at the same site, in the state $%
|ig\sigma >$ or an excited state of $H_i$ of symmetry $b_{1g}$ and energy $%
\sim 4$ above $|ig\sigma >$. Similarly for symmetry $B_{1g}$, there is a
small occupation of non-bonding $a_{1g}$ one-particle states of $H_i$ at
energy 4.30 above $|ig\sigma >$.

The position of the Raman $A_{2g}$ peak (0.18 below the band edge)
was calculated in a similar way. The
main difference is that the holes move in the two bands of $p_\pi $ states and that
there is a stronger effective on-site attraction.
As a consequence, the binding
energy of this exciton, with respect to the $p_\pi $ bands, is larger
(see Fig. 2)
and the wave function is more localized than for the previous solutions (
see Fig. 3).
For symmetry $B_{2g}$, the bound state falls above
the gap, inside the ZRS band. For our results to be compatible with the
Raman experiments \cite{liu} and the most accepted physical picture of the
hole-doped cuprates, it is necessary that while the lowest ZRS lies lower in
energy than the states containing $p_\pi $ orbitals, the $A_{2g}$ exciton
constructed with the latter lies inside the charge-transfer gap.
This is the case for reasonable parameters, like the ones we have taken.
 The binding energy of the $A_{1g},B_{1g}$ and $E_u$ excitons
is only sensitive to $U_{pd}$ and the shape of the quasiparticle band.
At least in the range $0.4 < U_{pd} < 1.2$ both $B_{1g}$ and $E_u$ excitons
bind and they lie less than $\sim 0.2$ below the gap edge, in agreement with
experiments in different compounds \cite{liu}. While
the position of the $A_{2g}$ peak is more sensitive to variations
in $\Delta_{xy},\Delta_\pi$ or $t_{\pi a}$, a different qualitative
explanation of it with 
parameters consistent with those given in the literature 
\cite{hyb,gra,ann,mat} is unlikely.

The intensity of the optical conductivity and Raman spectra is determined by
matrix elements of the kinetic energy in a given direction $T_x,T_y$, and of
the current operator ${\bf j}$ \cite{wag,kle,sha}. This operator can be
constructed from $H_{6b}$ in a standard way \cite{wag} and is linear in the
hopping terms. An analysis based on the form of the above mentioned
operators and the eigenstates of $H_i$, allows to infer the character of the
expected observable charge excitations of the insulating system. For example
the optical absorption of light with vector potential ${\bf A}\parallel 
\stackrel{\wedge }{\bf x}$ and finite frequency $\omega $ is proportional to 
$\sum_e\left| \left\langle e\mid j_x\mid g\right\rangle \right| ^2\delta
(\omega +E_g-E_e),$\ where $\left| g\right\rangle $ is the ground state and $%
\left| e\right\rangle $ are excited states \cite{wag}. The ground state of
the insulating system at each cell is given by $\left| ig\sigma
\right\rangle $, which corresponds mainly to a hole in a $%
3d_{x^2-y^2}$ orbital (a Cu$^{+2}$ state in agreement with spectroscopy
measurements \cite{tak}). The action of $j_x$ over this state gives $%
-i(a/2)t_{pd}(p_{i+a\stackrel{\wedge }{\bf x}/2,\sigma }^{\dagger }+p_{i-a%
\stackrel{\wedge }{\bf x}/2,\sigma }^{\dagger })\left| 0\right\rangle $. The
result has no overlap with states with one hole in $H_i$, but a significant
overlap with states with no holes at cell $i$ and two holes in a ZRS at a NN
cell $|(i\pm a\stackrel{\wedge }{\bf x})g2\rangle $. Thus, the lowest
optical excitation corresponds to energies near the charge-transfer gap, but
slightly below it because of the attraction between the electron left at
site $i$ and the additional hole at a NN site.

The dominant (resonant) contribution to the Raman intensity for incident
photon energy $\omega _i$, polarization $\stackrel{\wedge }{\alpha }$, and
the polarization $\stackrel{\wedge }{\beta }$ of the scattered light is
proportional to \cite{kle,sha}: 
\[
I\sim |\sum_e\frac{\langle f|j_\beta |e\rangle \langle e|j_\alpha |g\rangle 
}{E_e-E_g-\omega _i}|^2
\]
where $|f\rangle $ is the final state and $|e\rangle $ are excited
intermediate states. For experimental \cite{liu} values of  $\omega _i$  and
our results for the cell Hamiltonian, one can see that the excited states
with energies $E_e\sim E_g+\omega _i$ can be those with an empty site
(without holes) and another site with two holes: one of which is a linear
combination of a $3d_{x^2-y^2}$ hole and an O $b_{1g}$ Wannier function, and
the other can be an O $p_\pi $ hole (of symmetry $a_{2g}$ or $b_{2g}$, see Fig.
1), or a non-bonding O hole. The movement of the empty site distorts the
spin background, but, neglecting the dispersion of $E_e$ around the local
eigenenergies (eigenvalues of $H_i$) of the three two-hole states mentioned
above, the intermediate spin configurations can be summed, and only the spin
configurations of $|g\rangle $ and $|f\rangle $ enter in the expression for
the intensity. To take into account correctly these background spin
configurations, we have calculated the intensities from the exact
wavefunctions $|g\rangle $ and $|f\rangle $ in a 4x4 cluster, in the
Hilbert subspace of the corresponding irreducible representations.
We obtain that if
the incident frequency is near the energy required to excite an O $\pi $
hole of symmetry $a_{2g}$ ($b_{2g}$) then
$I_{A_{2g}}/I_{B_{1g}}=8.5 ~(15.7)$.

The origin of the Raman excitations is similar to that described in section
IV C. of Ref \cite{liu}. For symmetry $B_{1g}$ the O $p_\pi $ holes can also
take part in the intermediate states.

In summary, we have calculated the excitation energy and physical nature of
dipole active $E_u$ and novel Raman excitations of symmetries $\ B_{1g}$ and 
$A_{2g}$ observed experimentally in insulating cuprates. The $A_{1g},B_{1g}$
and $E_u$ peaks, as well as the low-energy excitations of the doped systems
can be well described by an effective generalized one-band model. The $A_{2g}
$ peak corresponds to a strongly bound exciton of $d_{xy}$ and $p_\pi $
holes. However, the amount of these holes in slightly doped systems is
very small and does not affect the ground-state properties.
For reasonable values of the parameters our calculated energies
agree with the experimentally observed ones for the optical absorption peak 
\cite{fal,per} and  Raman excitations \cite{liu}.

We thank M. V. Klein and D. Salamon for discussions about the experimental
data. Four of us (M.E.S., C.D.B., E.R.G., and F. L.) are supported by
CONICET. A.A.A. is partially supported by CONICET. Partial support from
Fundaci\'{o}n Antorchas under grant 13016/1 is gratefully acknowledged.

\figure {\noindent FIGURE 1:\\Representation of the Wannier functions at a
given Cu site ( denoted by a cross) constructed from the O $2p_\sigma$
orbitals (top) and $2p_\pi$ ones (bottom).}

\figure {\noindent FIGURE 2:\\Scheme of the one-hole spectral density of the
extended one-band Hubbard model (right, full line), the quasiparticle band
for only one added hole or electron ( right, dashed line), the spectral
density of the $d_{xy}$ and $p_\pi$ states (left) and the position of the different
charge-transfer excitations (straight lines).}

\figure {\noindent FIGURE 3:\\Amplitude of the wave function of the excitons
of different symmetry as a function of the relative coordinate between the
hole and electron. The radius of each circle is proportional to the largest
amplitude of the Hubbard one-hole creation operator of the corresponding
cell (like $\left| ig2\right\rangle \left\langle ig\sigma \right| $ for
symmetry $E_u$ at any Cu site). Full and empty circles correspond to
opposite signs. Crosses correspond to zero amplitude. For $A_{1g}$ we have
used $V_1=0.26$ to bind it.}

\end{document}